\documentclass{vgtc}                          % final (conference style)
\ifpdf%                                % if we use pdflatex
  \pdfoutput=1\relax                   % create PDFs from pdfLaTeX
  \usepackage{graphicx}                % allow us to embed graphics files
  \DeclareGraphicsExtensions{.pdf,.png,.jpg,.jpeg} % for pdflatex we expect .pdf, .png, or .jpg files
\else%                                 % else we use pure latex
  \usepackage{graphicx}                % allow us to embed graphics files
  \DeclareGraphicsExtensions{.eps}     % for pure latex we expect eps files
\fi%

\graphicspath{{figures/}{pictures/}{images/}{./}} % where to search for the images

\usepackage{balance}
\usepackage{microtype}                 % use micro-typography (slightly more compact, better to read)
\PassOptionsToPackage{warn}{textcomp}  % to address font issues with \textrightarrow
\usepackage{textcomp}                  % use better special symbols
\usepackage{mathptmx}                  % use matching math font
\usepackage{times}                     % we use Times as the main font
\usepackage{cite}                      % needed to automatically sort the references
\usepackage{tabu}                      % only used for the table example
\usepackage{booktabs}                  % only used for the table example
\usepackage{gensymb}

\usepackage{booktabs, multirow} % for borders and merged ranges
\usepackage{soul}% for underlines
\usepackage[table]{xcolor} % for cell colors
\usepackage{changepage,threeparttable} % for wide tables

\onlineid{0}

\vgtccategory{Research}

\vgtcinsertpkg

\title{SURVIVRS: Surround Video-Based Virtual Reality for Surgery Guidance}

\author{
        Amani Taweel\thanks{e-mail: ataweel@uncg.edu}\\ %
         \scriptsize Dept. of Computer Science\\\scriptsize UNC Greensboro %
        \and Joaquim Jorge\thanks{e-mail: jorgej@tecnico.ulisboa.pt}\\ %
         \scriptsize IST - ULisboa, INESC-ID %
        \and Anderson Maciel\thanks{e-mail: anderson.maciel@tecnico.ulisboa.pt}\\ %
         \scriptsize IST - ULisboa, INESC-ID %
         \vspace{8pt}
        \and João Ricardo Nickenig Vissoci\thanks{e-mail: jnv4@duke.edu}\\ %
         \scriptsize Duke Global Health Institute\\\scriptsize Duke University%
        \and Regis Kopper\thanks{e-mail: kopper@uncg.edu}\\ %
        \scriptsize Dept. of Computer Science\\\scriptsize UNC Greensboro %
}

\teaser{
  \includegraphics[height=2in]{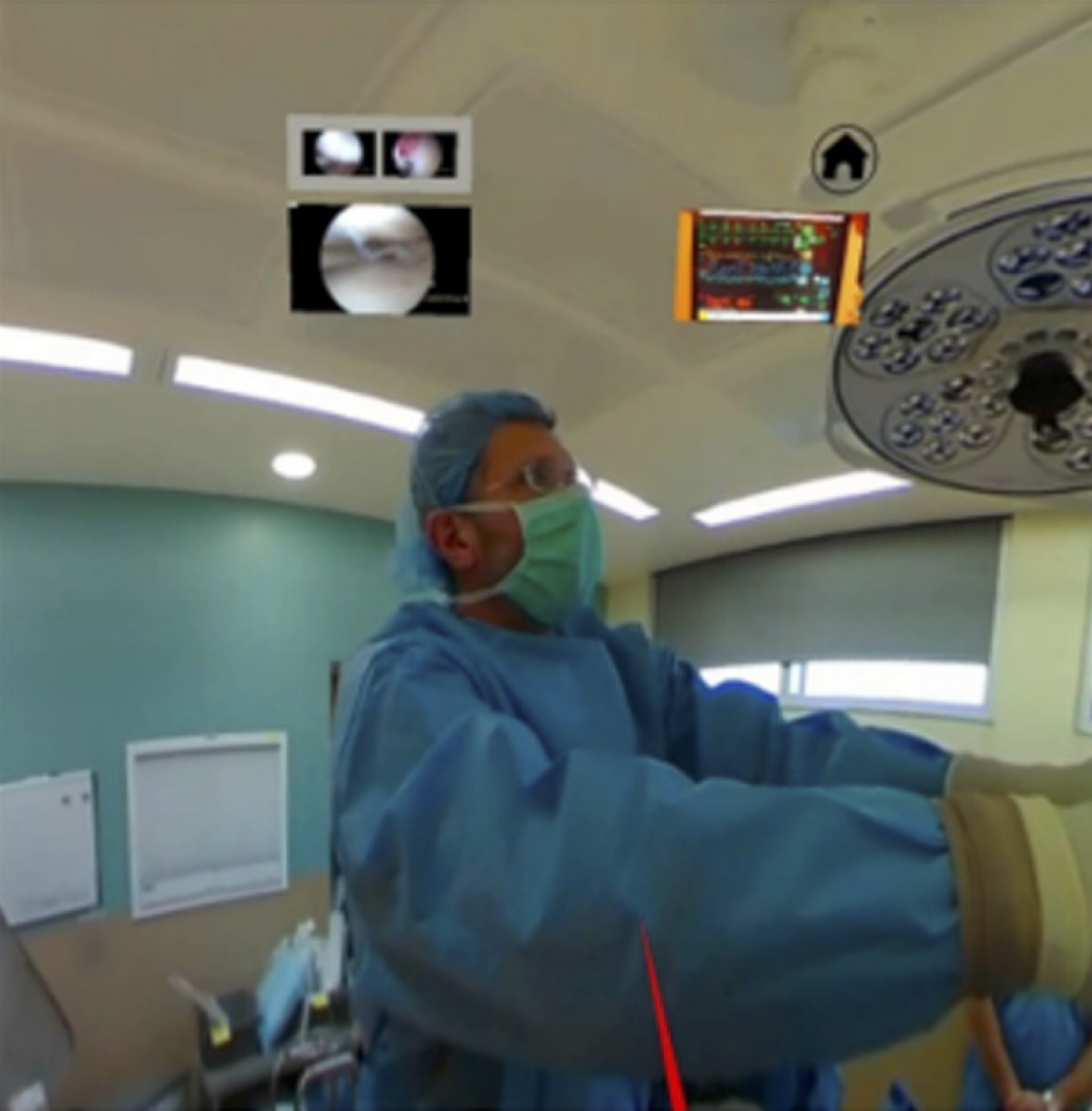}
  \includegraphics[height=2in]{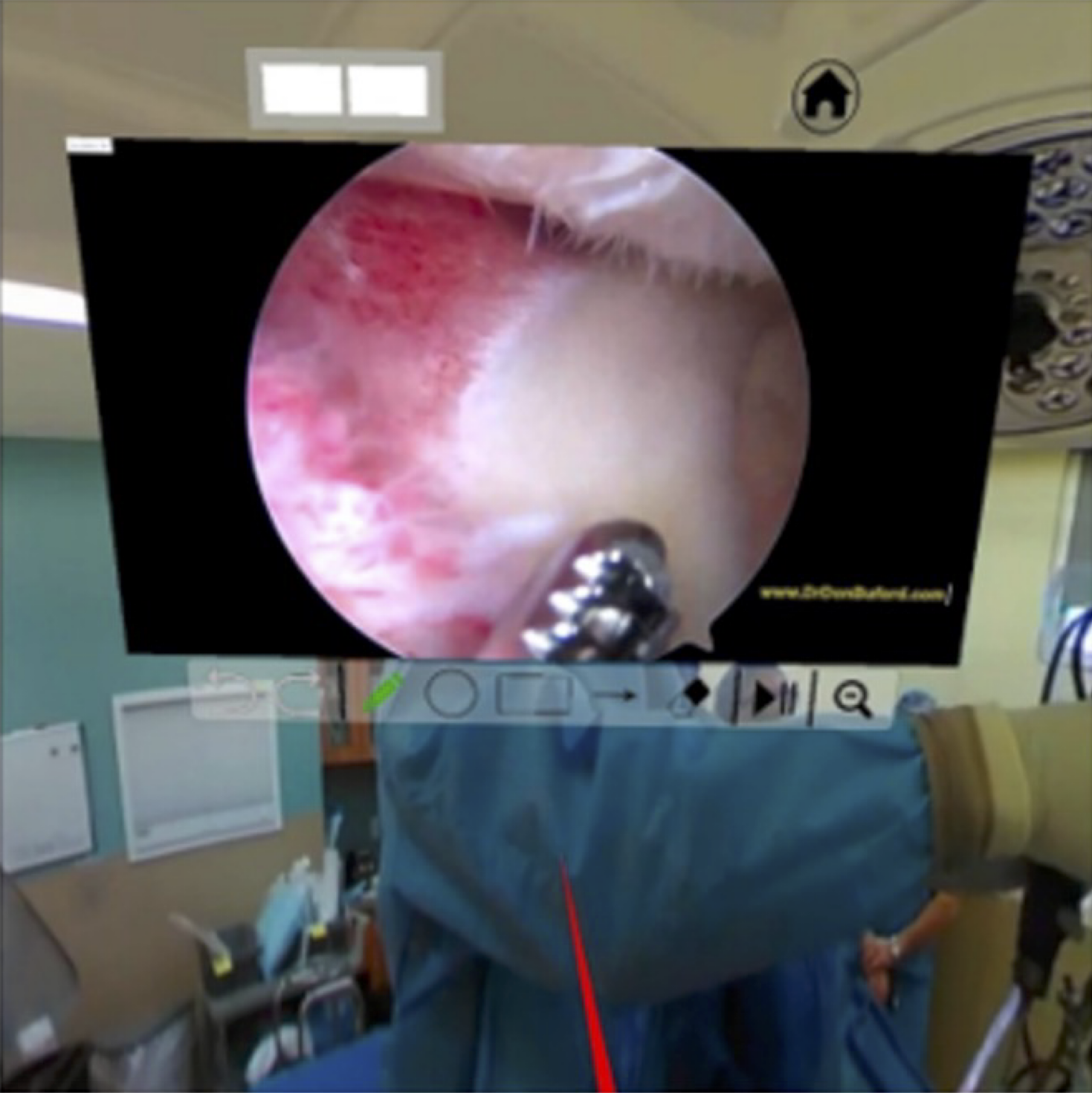}
  \includegraphics[height=2in]{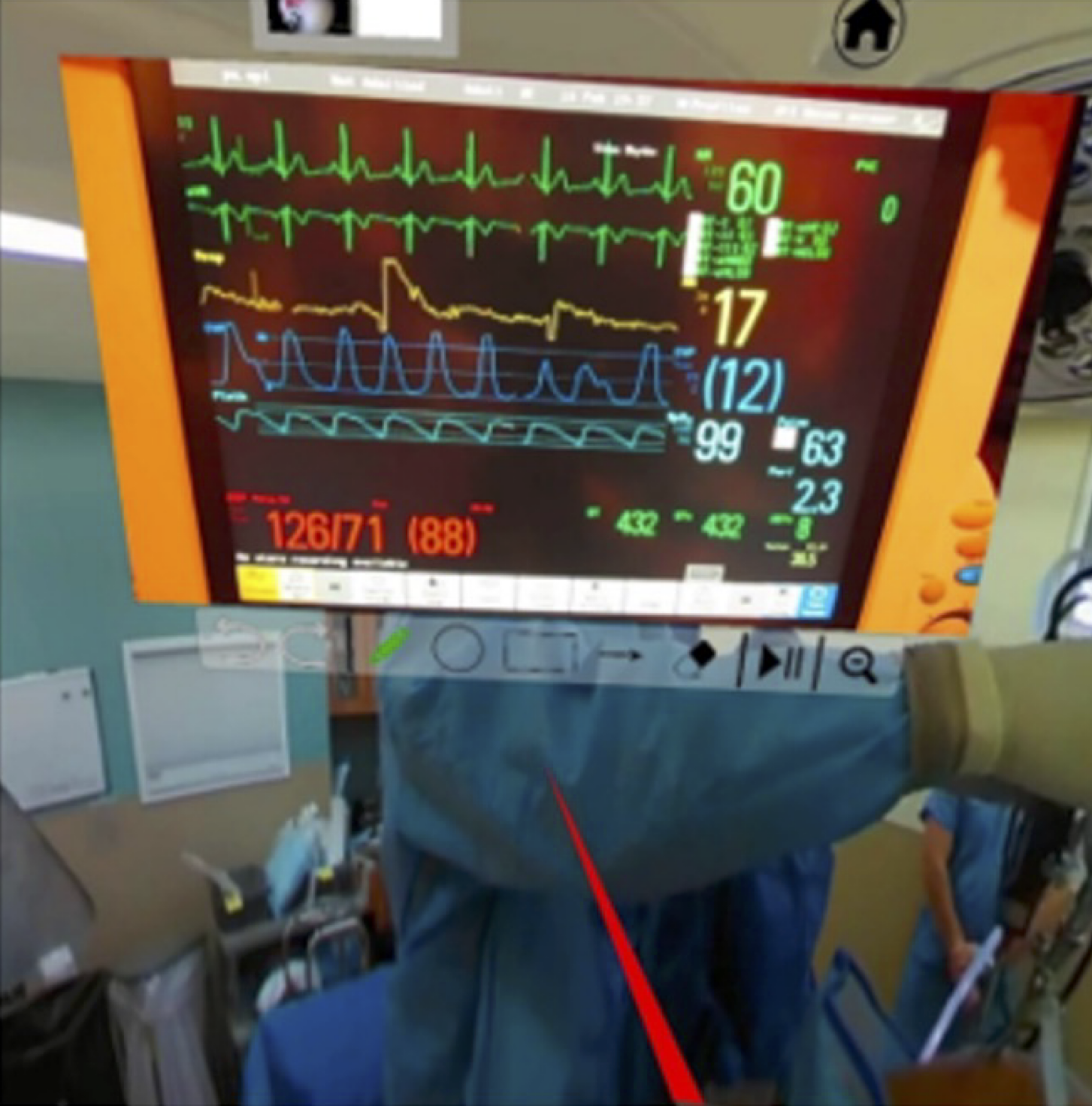}
  \caption{The SURVIVRS remote surgery guidance tool. (left) Immersive visualization of the surgery room offers a telepresence experience to the remote surgeon. (center) The local site of the surgery can be visualized in detail, where annotations can be sent to the surgery room. (right) Similarly, a live view of the patient vitals monitor is detailed to which the surgeon can make annotations.}\label{fig:teaser}
}

\abstract{There is a strong demand for virtual reality (VR) to bring quality healthcare to underserved populations. This paper addresses this need with the design and prototype of SURVIVRS: Surround Video-Based Virtual Reality for Surgery Guidance. SURVIVRS allows a remote specialist to guide a local surgery team through a virtual reality (VR) telepresence interface. SURVIVRS is motivated by a need for medical expertise in remote and hard-to-reach areas, such as low-to-middle-income countries (LMICs).
The remote surgeon interface allows the live observation of a procedure and combines 3D user interface annotation and communication tools on streams of the surgical site and the patient vitals monitor. SURVIVRS also supports debriefing and educational experiences by offering the ability for users to watch recorded surgeries from the point of view of the remote expert. The main contributions of this work are: the feasibility demonstration of the SURVIVRS system through a rigorous 3D user interface design process; the implementation of a prototype application that realizes the proposed design; and a usability evaluation of SURVIVRS showing that the tool was highly favored by users from the general population. The paper discusses the next steps in this line of research aimed at more equitable and diverse access to healthcare.

} % end of abstract

\CCScatlist{
  \CCScatTwelve{Human-centered computing}{Human computer interaction (HCI)}{Interaction paradigms}{Virtual reality}
}

\begin{document}

% the only exception to this rule is the \firstsection command
\firstsection{Introduction}

\maketitle

Many developing countries worldwide show an unmet demand for medical professionals with specific specialties. As an example, in 2007 there were approximately only 27 neurosurgeons available in East Africa, across seven countries with a total population of 270 million, placing the ratio of neurosurgeons to residents at 1:10 million~\cite{fuller2016building}.\footnote{As a means of comparison, the ratio of neurosurgeons to residents in the United States was approximately 1:65,580 in 2007~\cite{fuller2016building}.} Virtual reality (VR) technology has the potential to reduce such gaps in healthcare access.

The major problem this research proposes to address is the lack of specialized surgeons in hard-to-reach areas at an accessible cost. Specifically, this project offers a means to enable collaboration between surgeons who are geographically apart. This collaboration happens through live communication via immersive surround video streams in virtual reality. This allows remotely located surgeons to guide and give instructions to the surgery room on short notice without the need to travel or incur additional costs.

Our proposed solution is SURVIVRS: Surround Video-Based Virtual Reality for Surgery Guidance, an immersive virtual reality telepresence tool for remote surgeons to guide medical procedures in real time. Figure \ref{fig:teaser} shows the main components of SURVIVRS.

SURVIVRS contains three main display components: a live 360\degree~stream of the surgery room, from the vantage point of a locally mounted 360\degree~camera (henceforth referred to as \textbf{\textit{360\degree~view}}), a live high-resolution stream of a spot camera aimed at the local site of the surgery (henceforth referred to as \textbf{\textit{site view}}), and a live visualization of the patient vitals monitor (henceforth referred to as \textbf{\textit{vitals view}}). We posit that, for the remote expert surgeon to offer fully informed guidance to a less experienced surgery team, it is important that they have a contextual view of the surgery theater. As long as the remote surgeon is connected to SURVIVRS, they maintain their spatial context and orientation through the 360\degree~view that provides a first-person immersive video stream rendered to a VR head-mounted display (HMD). The site view and the vitals view are always displayed near the upper corners of the remote surgeon's view and can be zoommed in on demand. 

%Everything in actual surgery must be accurate, rigorous, and punctilious. So, in some surgeries, more than virtual reality and 3D imaging is needed to show the necessary details. To address this challenge, this research proposes using a 360\degree~camera inside the surgery room to offer an immersive first-person view of the surgery room in real-time, including the patient, medical staff, and equipment. To allow for more accuracy and perception of the site of the surgery and the patient's state, the proposed tool uses two additional sources of information. Involving the real-time image of a camera pointed directly at the surgery site, and the second lives feed of the patient vitals monitor, showing data such as blood pressure and heart rate. 

In addition to live surgery guidance, SURVIVRS also serves an educational benefit.  Traditionally, surgery had been taught through the ``See One, Do One, Teach One'' method, where trainees perform procedures after observing similar ones. Although this method has been criticized as unsafe to patients, its core process remains unchanged~\cite{kotsis2013application}. SURVIVRS can potentially improve the ``See One'' stage of surgical training by playing back surgeries to trainees. The replay includes all tools available to the remote surgeon during the live procedure, including the 360\degree~view, the site view, and the vitals view. This way, students can review a surgery, even adding annotations on any necessary parts of the stream.

This project used several tools for its development. It uses the Unity game engine to provide a VR application that offers immersive telepresence to a remote user (the expert surgeon) with tools that allow the guidance of a procedure. For this application to succeed, it needs to ensure a friendly, intuitive, and usable immersive virtual environment on the remote surgeon's side. To that end, an annotation system with varied input options was developed to allow the surgery team to visualize guidance cues during the surgery precisely. Another goal of this research was to create a live streaming link between the VR application (surgeon's side) and a web app (simulating the surgery room side) with high resolution, low cost, and a clear audio connection. The final goal was to make a web application that emulates the surgery room side easy to use with options to change camera sources, resolutions, and tools to record the call. 

In summary, the main contribution of this work is the design of SURVIVRS, a VR interface that connects a specialist surgeon to a surgery room, allowing them to guide a surgery team through the course of a procedure. This research is motivated mainly by the prospect of SURVIVRS enabling efficient communication between surgical teams in LMICs and other hard-to-reach areas and experienced specialists in developed countries to ultimately improve access to critical care and outcomes in remote locations.

%This paper is organized as follows: section \ref{sec:related} provides the related literature to this research and contrasts it with our proposed approach. Section \ref{sec:methodology} presents the tool design in detail. Section \ref{sec:evaluation} contains details about the evaluation of the proposed tool’s user interface (UI), its results, and recommendations for improvements. The paper concludes with section \ref{sec:conclusion}, where takeaways from this work are discussed, and suggestions for future work are presented.

\section{Related Work}
\label{sec:related}

%This section reviews work related on virtual reality use in healthcare, the meaning and implications of interacting with 360\degree~video, and covers related research that use virtual reality in remote guidance tasks.

\subsection{Virtual Reality in Healthcare}

Virtual reality is a way to connect humans to computers by using HMDs to simulate virtual and interactive environments. VR provides an immensely realistic and controllable environment for users to interact across many application areas. These VR features have been used in experiences and clinical settings for over two decades~\cite{rizzo2017clinical}. One of the first real-world applications of VR was in 1993 to treat mental health disorders. Then VR applications were developed to handle specific phobias~\cite{aziz2018virtual}.

McCloy and Stone~\cite{mccloy2001virtual} argue that VR simulators are suited for general surgery, while Augmented Reality (AR) is suited to guide neurosurgery. SURVIVRS offers a remote AR-like interface by giving the expert surgeon a real-time view of the live surgery room. 

Our research aims to employ VR technology to allow communication between remote medical specialists and non-specialist surgeons in a surgery room with the goal of providing instructions for a more successful surgical outcome. Using a VR application, we propose a prototype tool that, in the future, could work over a connection between a doctor from a developed country and a surgery room in an LMIC. Also, we offer the ability for a remotely guided procedure to be recorded, raising the opportunity for medical students to replay the VR view of the remote surgeon to learn and review the surgery.

\subsection{Interaction with Immersive 360\degree~Videos}

Technology of 360\degree~video uses omnidirectional camera systems and allows viewing with a first-person perspective. %While playing the video, users can interact with and move the video by mouse or by panning and tilting a mobile device to change the viewing degree of the video~\cite{snelson2020educational}. 
These experiences have been described as highly immersive and potentially engaging environments surrounded by a total sense of existence~\cite{argyriou2016engaging}.

\subsection{Remote Collaboration through Virtual Reality}

%This section reviews some research that uses VR with remote guidance.

Connecting the world and sharing real-time experiences between people who may be geographically distributed is now an urgent need. Specifically, there is a demand for remote guidance, where a remotely situated expert assists their team on the other end of some procedure, such as a repair, maintenance, surgery, or communication task. This type of collaboration is required in many fields, such as education and healthcare. 

Huang et al.~\cite{huang2013handsin3d} developed a 3D system called HandsIn3D based on 3D virtual immersed space and 3D hand gestures. This project connects a worker and a helper through 3D stereoscopic cameras that offer a visualization of the work site by the helper and augments the worksite by rendering representations of the helper's hands.

Comparing the HandsIn3D project with our own research, they are based on common principles of remote live video-based guidance, but the implementation and goals are entirely different. While HandsIn3D uses a 3D camera to get the other side, SURVIVRS uses a 360\degree~camera to film the surgery room entirely. Moreover, instead of creating virtual hands on the work side to help people, our work offers an annotation system to allow a remote expert to live annotate and take screenshots. Finally, the main goal of the HandsIn3D project was to remotely help distant workers to complete manual tasks, whereas our project targets live remote guidance to a surgical procedure.

\section{Design and Implementation}
\label{sec:methodology}

%This section describes the progress of user interface design step by step to develop the proof of concept of the interface and ensure interaction efficiency under different design options. Based on the iterative design process, this project has selected the final interface design to meet the users' expectations as much as possible.% This research has already tested the 3D interfaces with volunteers. The experiment and its results are shown in section \ref{sec:evaluation}.

%Finding the best UI design for the project the first time is unexpected to occur. That is why the UI design needs to be intrinsically open, iterative, and incomplete throughout the project’s execution to allow modification at any time in the process~\cite{coyette2007multi}. This means UI design needs to consider the progress of appropriate techniques, such as prototypes and final designs, which are covered in the coming subsections. 

%\subsection{Early Prototyping}

We used an iterative design process to propose and refine the design of SURVIVRS.
%This project depends on the prototype fidelity concept since many rapid changes are needed for UI prototyping. Prototype fidelity means the similarity between the final coded user interface and the prototyped user interface. There are three types of fidelity prototypes: high-fidelity, when the prototype is the closest to the final design in layout and navigation. The second type is a low-fidelity prototype when the prototype represents part of the final UI, not all of its details. The third one is between low and high fidelity and is called medium fidelity~\cite{coyette2007multi}. 
%
%This research uses the low-fidelity prototype to design screens on both surgeon and surgery room sides. On the surgeon side, which represents the VR project side, there is a 360\degree~view for the user and two other regular videos to interact with. The progress of getting an exemplary user interface with efficient use takes a lot of time and steps to test and redo. One of the earlier designs is shown in Figure~\ref{fig:prototype}(b) the live streaming main menu missed the screenshots list, Figure~\ref{fig:prototype}(d) the drawing tools menu is placed over the 2D videos, now it is under the video, and the buttons were not all in the same place, some of them were in the top corners of the video.
Early on, we used low-fidelity prototyping to experiment with a variety of interface ideas. This allowed the research team to openly discuss ideas without loss of code-based work.

After finishing the iterative design process, we conducted several steps to ensure the final usability of the tool. Those steps involved developing the user interfaces by implementing code to the prototypes, testing the design by running and experiencing it, and finally evaluating the usability of the tool by study volunteers. Figure~\ref{fig:teaser} shows screenshots of the SURVIVRS prototype remote surgeon interface used in the usability evaluation.

%  \begin{figure}[htb]
%  \centering% avoid the use of \begin{center}...\end{center} and use \centering instead (more compact)
%  \includegraphics[width=\linewidth]{pictures/main_UI.png}
%  \caption{Surgeon Side Screenshots. (a) Main Menu Screenshot. (b) Main User Screen. (c) The specific Surgery Site Video Screen. (d) The Patient’s Data Monitor Screen.}
%  \label{fig:main_ui}
% \end{figure}

\begin{figure}[htb]
 \centering% avoid the use of \begin{center}...\end{center} and use \centering instead (more compact)
 \includegraphics[width=\linewidth]{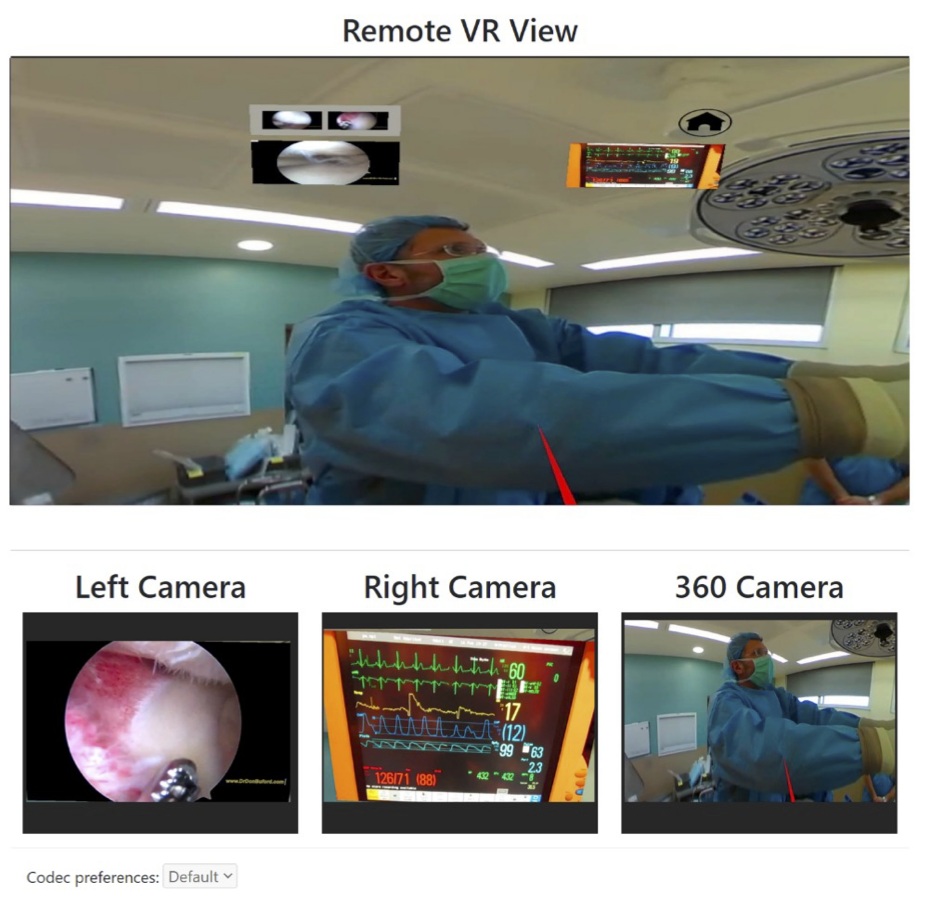}
 \caption{Screenshot of the web application simulating the surgery room. A stream of the remote surgeon's viewpoint is shown at the top, while raw streams of each camera feed is displayed at the bottom.}
 \label{fig:webapp}
\end{figure}

As the focus of this project is the implementation of the remote surgeon's guidance user interface, we simulated the surgery room through a web application (Figure \ref{fig:webapp}) that streams the view of the remote surgeon with virtual cameras and screen-capture software.

\subsection{Software Tools}

%This section describes all software and hardware tools used to design and implement the Virtual Reality Remote Surgery Guidance (VRSG) system. The following section explains in detail how these tools are connected to make an efficient working system.

%\subsubsection{Software Overview}

%This subsection discusses the software part and all programs that are used to build the SURVIVRS system as follows:

%\textbf{\textit{Unity Game Engine}}

%Unity is a real-time cross-platform tool that consists of rendering, a game engine (including physics), a user interface (Unity Editor), and an integrated development environment (IDE). Unity is known and mainly used in the 3D development, gaming, and film industries. These games and applications are developed for different Augmented Reality (AR) or Virtual Reality (VR) platforms~\cite{juliani2018unity}.

We implemented the SURVIVRS prototype in Unity\footnote{https://www.unity.com} (version 2020.3.33f1) with OpenXR\footnote{https://www.khronos.org/openxr}. The VR application uses the Unity Video player as a broadcaster for the 360\degree~video which is rendered as a texture onto the scene's Skybox. Unity Render Streaming (URS) was used to live stream the VR view of the remote surgeon to the simulated surgery room web application. Finally, OBS Studio was used as a virtual camera to stream the contents of the live 360\degree~video footage captured by the 360\degree~camera.

\subsection{Hardware}

% This subsection describes the hardware tools used to build the SURVIVRS prototype. These tools are listed as follows:
% \begin{itemize}

We used the Oculus Meta Quest 2 with two tracked controllers as the VR platform for the remote surgeon. Live streaming of the 360\degree~video was achieved by an Insta360 Pro camera\footnote{https://www.insta360.com/product/insta360-pro}. For this prototype, we used two spot cameras to capture the image of the local surgery site, and a view of the patient vitals monitor. For future implementations, we expect to render a digital vitals monitor using the patient's data, rather than a video feed, which is lower quality and unnecessarily takes up bandwidth.

The surgery room simulation is run in a server PC with WiFi to communicate with the VR application. All three cameras and a monitor are connected to this PC, which displays the surgery room simulation web application.

% \item Virtual reality headset and controllers: this project finds that Oculus Meta Quest 2 is the most appropriate for the requirements. The remote surgeon wears the headset to communicate with the surgery room via the VR application. As well as using a VR controller to zoom in/out the two videos and draw annotations inside the VR application. Figure~\ref{fig:action}(a) shows a surgeon wearing the Oculus Meta Quest 2.
% \item 360\degree~camera: this project uses an Insta360 Pro camera to capture a video of the whole surgery room and send it to the Insta360 Pro application.

% \item Two 2D cameras: these two cameras are used in the surgery room where the first one is used to spot a specific surgery site of the patient, and the second camera is to show the patient data monitor. See Figure~\ref{fig:action}(b) of the surgeon's view to note that the top-left video is from camera 1 and shows the specific surgery site, and the top-right video is from camera 2 for the patient data monitor.
% \item Screen (monitor) in the surgery room to see the remote surgeon’s view to be able to see the annotations which are made by the surgeon. Figure~\ref{fig:action}shows the surgery room with the screen displaying the specific surgery site video zoomed out and annotated by the remote surgeon. 
% \item The surgery room's PC (computer/server) is connected to the WiFi to communicate with the VR application (remote doctor). And all three cameras and the screen (monitor) are connected to this PC.

% \end{itemize}

\subsection{System Design}

%This section views the project’s idea and system structure in detail. Also, it describes how we put together all software and hardware equipment from the previous section to shape the final implementation. Finally, this section highlights the project’s modes of live connection mode, view recorded video mode, and view the Annotation system.

\begin{figure}[!b]
 \centering% avoid the use of \begin{center}...\end{center} and use \centering instead (more compact)
 \includegraphics[width=\linewidth]{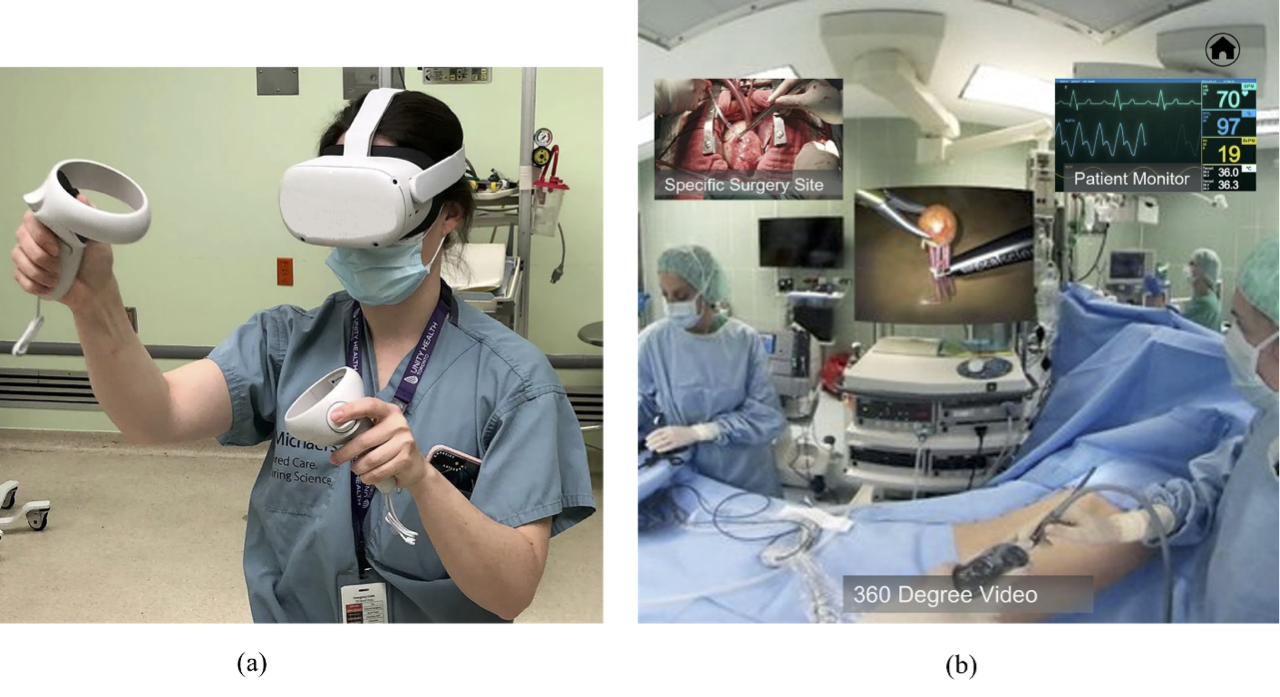}
 \caption{VR Application on The Surgeon Side. (a) The Idealized Image of a Doctor Using the VR Application. (b) VR Application Interface on the Remote Surgeon Side.}
 \label{fig:action}
\end{figure}

The primary purpose of this project is to offer a first step towards enabling a remote surgeon to guide a surgeons’ team in a remote area such as sub-Saharan Africa. %This project consists of two spaces: the first space is an immersive VR application for the remote surgeon, as shown in 
Figure~\ref{fig:action} shows a conceptual image of a doctor wearing the guiding a remote surgery (a) and their corresponding VR view (b).% This VR application streams three videos in the same interface: a 360\degree~video to provide an immersive view of the surgery room, and two other videos are arranged on the 360\degree~video: the top-left video for a specific surgery site video and the top-right video for the patient data monitor as in Figure~\ref{fig:action}(b). 
The surgeon can turn in any direction to see everything in the surgery room. As the user changes their view direction, the 2D heads-up UI follows their head movement in all directions after a 0.5-sec delay through a smooth animation.

% The second user interface, shown in Figure~\ref{fig:ideal_team}, represents the surgery room in a simplification that allows the emulation of the idealized complete project while maintaining the scope of the work focused on the remote surgeon user interface. There are several tools in the surgery room application: a 360 camera to record the whole room, two 2D cameras: one directly over the patient to record the specific site of the surgery, another showing the patient’s data monitor with data such as pulse and blood pressure, a computer/PC to connect all these cameras to it, and a screen/monitor to display the connection with the remote surgeon for the surgeon team.

%  \begin{figure}[htb]
%  \centering% avoid the use of \begin{center}...\end{center} and use \centering instead (more compact)
%  \includegraphics[width=\linewidth]{pictures/ideal_team.png}
%  \caption{Mock-up Surgery Room Side.}
%  \label{fig:ideal_team}
% \end{figure}

% Mock-up Image to Show the Idealized Use of Surgeon’s Team Looks at an Annotated Video from the Remote Surgeon.
 
\begin{figure}[!t]
 \centering% avoid the use of \begin{center}...\end{center} and use \centering instead (more compact)
 \includegraphics[width=\linewidth]{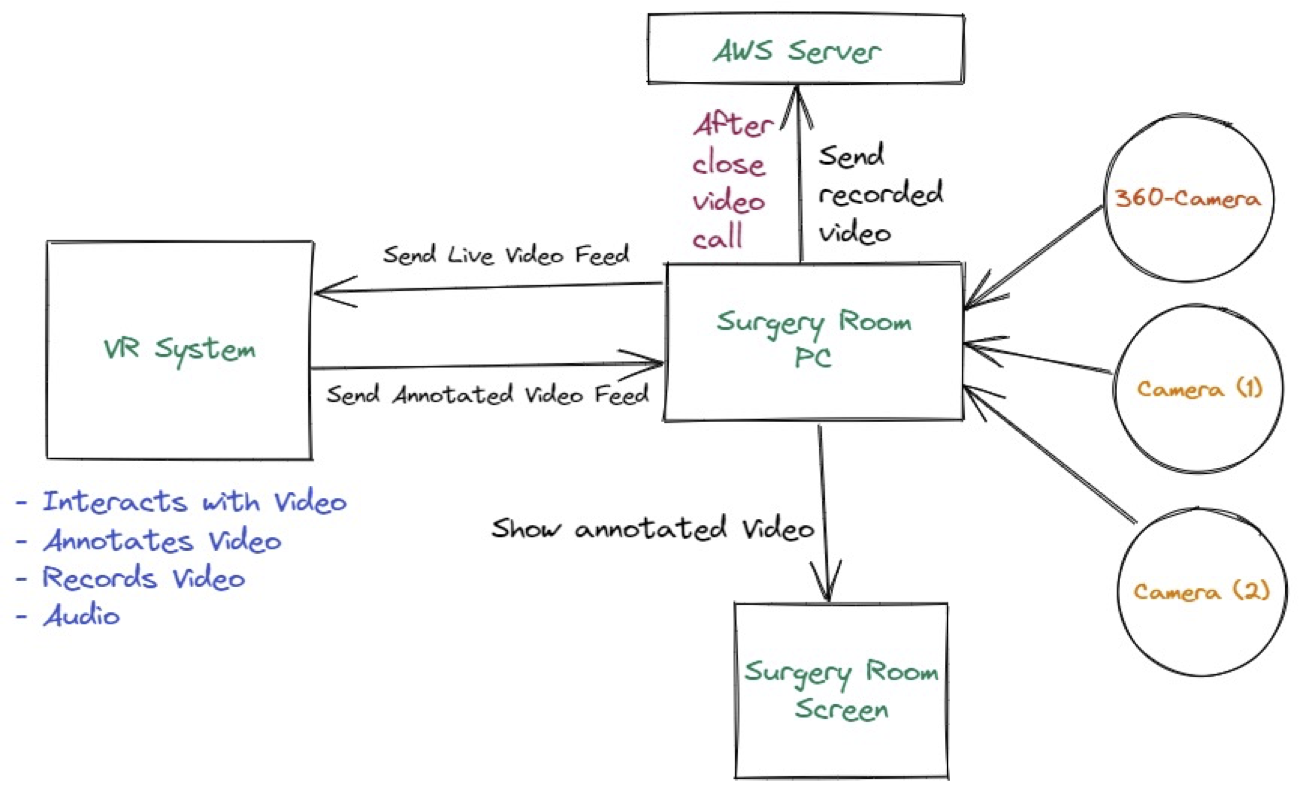}
 \caption{Information Flowchart.}
 \label{fig:flow}
\end{figure}

Figure~\ref{fig:flow} shows the information flowchart of the entire communication between the VR application and the simulated surgery room web application. The flow starts from any side by the connection ID to start streaming videos over a WebSocket after finishing the signaling between them. The surgery room web app streams the 360\degree~video, the local surgery site video, and tge patient data monitor video to the VR application on the remote surgeon side. On the VR application side, the specialist can interact with these videos by annotating them in the live stream. All these annotations as well as the three streams are shown on the surgery room screen to be viewed by all surgeons (Figure~\ref{fig:webapp}). A live two-way audio connection is also established. After closing the stream, a copy of the video call is saved on the server to be used in the ``view-recorded video'' mode in the VR application by medical students or other doctors in our idealized scenario.

\subsubsection{Annotation Interface}

\begin{figure}[htb]
 \centering% avoid the use of \begin{center}...\end{center} and use \centering instead (more compact)
 \includegraphics[width=\linewidth]{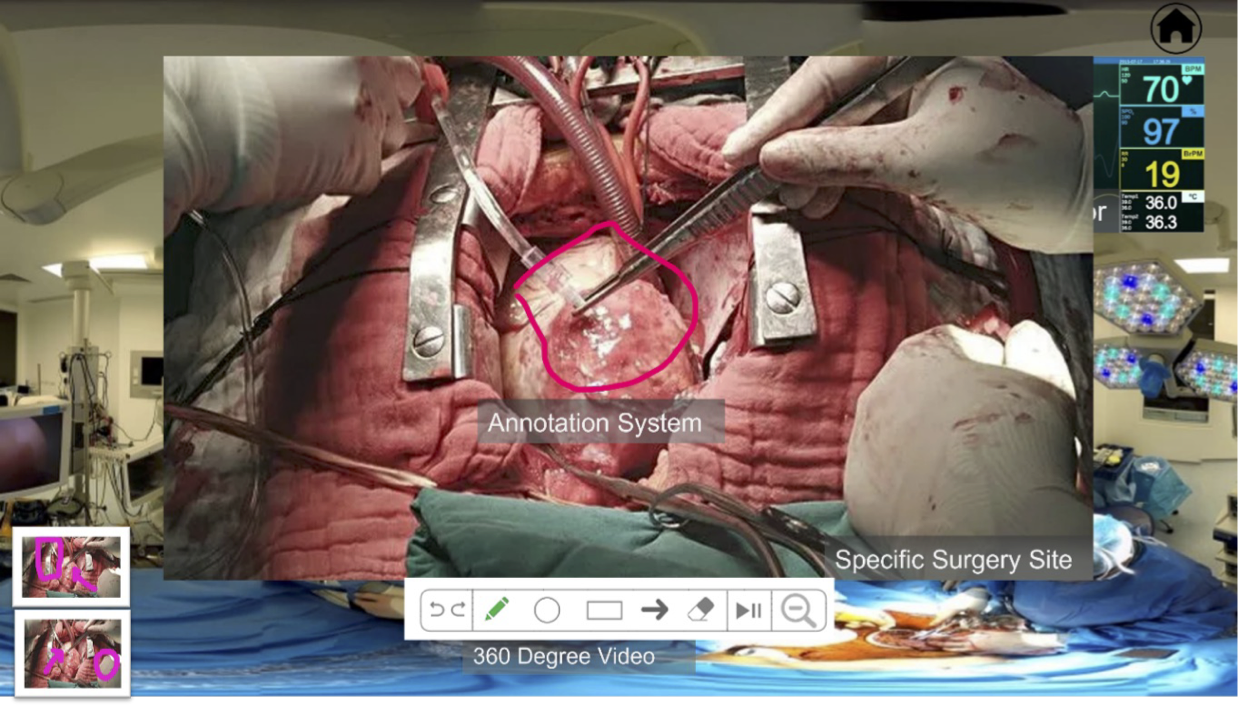}
 \caption{Annotation System.}
 \label{fig:annot}
\end{figure}

One of the most essential features of SURVIVRS is to give the remote surgeon the ability to guide the surgery by pointing to details and convey them to the surgery room through an annotation interface. The annotation system appears on the 2D streams when the surgeon points the controller to one of the 2D videos in the main interface and clicks the trigger buttin to zoom it in. After the video becomes larger, a menu of drawing tools is displayed in front of the surgeon in the VR application, as shown in Figure~\ref{fig:annot}. The menu has four annotation supporting tools: undo and redo; shapes (pencil--free drawing, oval, rectangle, arrow, and eraser); and a play/pause button which, in addition to playing and pausing the video, also takes a screenshot of the annotated 2D video (Figure~\ref{fig:annot} shows previously taken screenshots in the left-bottom corner of the UI. Whenever an annotation starts, the 2D video is automatically paused, so that image movement does not disturb the annotation target.

\section{Evaluation}
\label{sec:evaluation}

%The main goal of doing experiments is to test the efficiency and usability of the 3D user interface in the VR application (surgeon side) by using ``View recorded videos'' mode. This section describes the design and conduct of this project’s evaluation.

%All researchers took Institutional Review Board (IRB) training in human resources in the Collaborative Institutional Training Initiative (CITI Program) at UNCG. Then applied to the Cayuse Human Ethics IRB Application and got the approval to experiment with the Interactive Realities Lab (IRLab) at UNCG (IRB-FY23-128).

We performed a usability evaluation of the SURVIVRS prototype to understand how users from the general population react to the tool before we test it with less available medical doctors. Institutional Review Board approval was obtained prior to data collection. The evaluation was performed with the ``view recorded videos'' mode to allow the experience of a real surgery video and avoid technical issues such as latency that are not part of the user interface.

\subsection{Participants}

For this evaluation, six volunteers participated in the experiment, four had no vision issues, and two had strabismus. There were three females and three males, ages ranging from 18 to 44. Most of the participants (66.6\%) rated themselves as slightly familiar with VR, while the rest were either not familiar with VR at all (16.6\%) or moderately familiar with VR (16.6\%).

\subsection{Measures}

Overall perceived usability was measured with the System Usability Scale~\cite{brooke1996sus}. Presence was measured with the Slater-Usoh-Steed presence questionnaire~\cite{usoh2000using}. We also developed an overall perceptions questionnaire with specific questions about the design and functionality of SURVIVRS user interface.

\subsection{Procedure}

% This experiment was done at IRLab in the Petty Science Building at UNCG. The study was done with each participant at a scheduled time. Before starting the investigation, the participant had to sign a consent form and fill out a background questionnaire. And a study ID was assigned for each participant for any data collection instead of their personal information, so this experiment maintains confidentiality. 

After the initial greeting and signing the consent form, participants were given general instructions about the experiment. Then, participants wore the HMD and controllers to begin the experiment. At all times, during data collection, the experimenter observed through casting from the HMD to a PC. 

After ensuring that the participant got familiar with the VR HMD (Meta Quest 2), the experimenter asked the participant to complete a series of ten tasks that were meant to allow the exploration of all application features. No help or instruction on how to complete a task was given to the participants.

% The participants wore the VR headset to do the tasks. The headset was streaming to the experimenter's monitor/screen. After the experimenter had finished reading/explaining a task to the participant, the experimenter ran a stopwatch until the participant finished the task. Then wrote down the task number and spent time. This was done for every task until finished the ten tasks.

After finishing all tasks, participants were asked to complete the post-study questionnaires

% The tasks are ordered the same as the experiment, the participant didn’t have any training for these tasks before the experiment, and the instruction were oral:

% \begin{enumerate}
%     \item Select the third video from the recorded mode, then start.
%     \item Turn 90 degrees to the left to see the whole surgery room and ensure you can still see the videos.
%     \item Zoom in/out the patient data monitor video.
%     \item Pause the specific surgery video.
%     \item Go to the patient data monitor video and draw a circle over it.
%     \item Take a screenshot of what you annotated.
%     \item Act as if you are a doctor, indicate something important to other doctors in the specific surgery site video, then take a screenshot.
%     \item Display the last screenshot you have taken.
%     \item Open any video and draw anything, then delete what you draw.
%     \item Return to the main menu.
% \end{enumerate}

\section{Results}

% This section reports the evaluation results in four categories: completion time and success rates, system usability, perceived sense of being in the VR environment, and overall perceptions. All these results are based on the ten tasks and the post-study questionnaires mentioned in the previous section.

% \subsection{Completion Time}

% The first assessment tool to evaluate the ease of handling the user interface is by calculating the spent time in each task of the ten tasks by every user. Figure~\ref{fig:comptimes} represents tasks from one to ten on the X-axis and the average time users spend finishing each task in seconds (mean).

%  \begin{figure}[htb]
%  \centering% avoid the use of \begin{center}...\end{center} and use \centering instead (more compact)
%  \includegraphics[width=\linewidth]{pictures/completion_times.png}
%  \caption{Completion Time Mean Ratings.}
%  \label{fig:comptimes}
% \end{figure}

\subsection{System Usability}

% The second assessment evaluates the effectiveness of the user interface by examining its usability. This depends on the System Usability Scale questionnaire (SUS) questionnaire and the rates of the participants' answers to questions from Table 2 above. 

% To get the advantage of this scale, it generates a single number for each participant, representing a measure of the system's overall usability.

% To calculate this questionnaire in numbers, it is necessary to know its original scale is from strongly disagree = 1 to strongly agree = 5. The mean results of our experiment in the scale from 1 to 5 for each question is shown in Figure~\ref{fig:susmean}.

%  \begin{figure}[htb]
%  \centering% avoid the use of \begin{center}...\end{center} and use \centering instead (more compact)
%  \includegraphics[width=\linewidth]{pictures/SUS.png}
%  \caption{System Usability Scale Mean Ratings Per Question.}
%  \label{fig:susmean}
% \end{figure}

\begin{figure}[!t]
 \centering% avoid the use of \begin{center}...\end{center} and use \centering instead (more compact)
 \includegraphics[width=\linewidth]{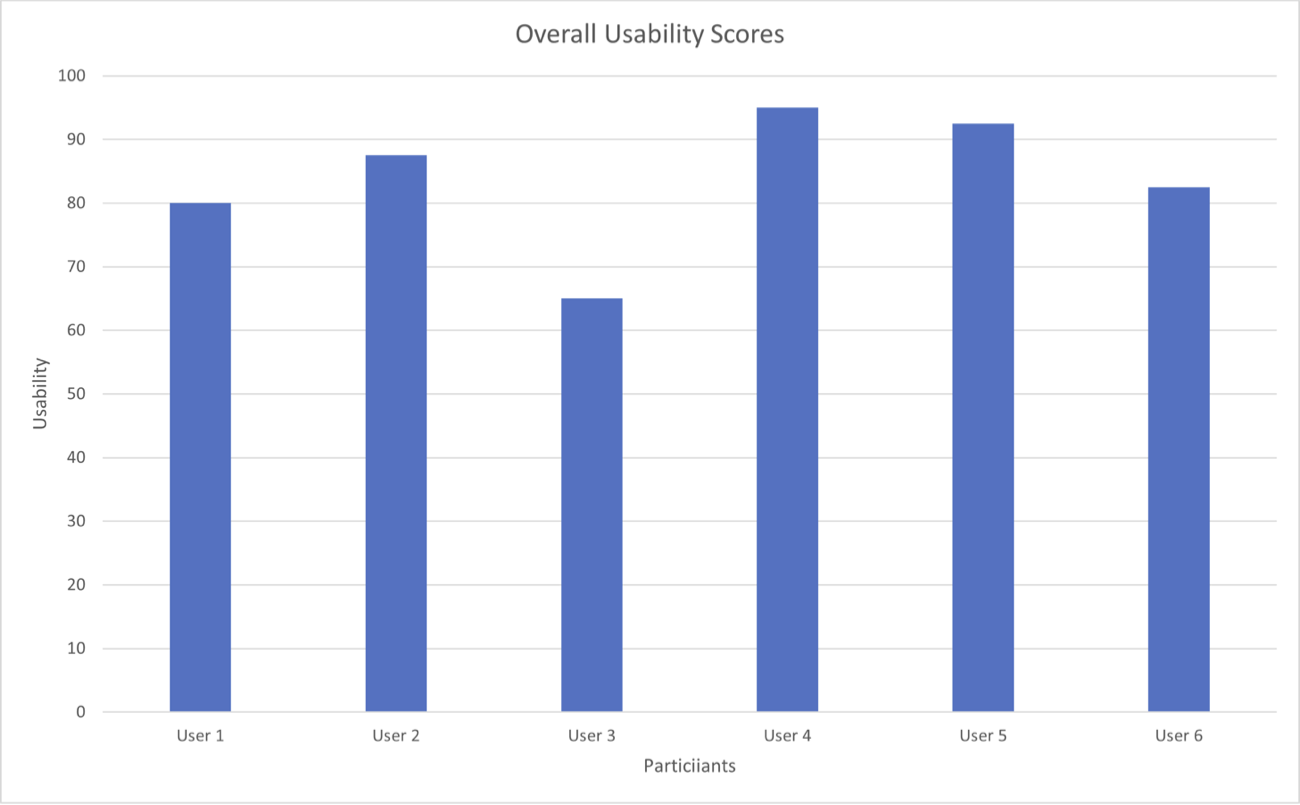}
 \caption{Overall Usability Scores for each Participant.}
 \label{fig:overalsus}
\end{figure}

Figure~\ref{fig:overalsus} shows the SUS Usability scores for each participant. We see that, for 5 out of the 6 participants, their scores are considerably higher than the 68-point threshold for average usability~\footnote{https://www.usability.gov/how-to-and-tools/methods/system-usability-scale.html}. 

\subsection{Presence}

Summary results for each question of the presence questionnaire are shown in Table~\ref{tab:respres}.

\begin{table}[!t]\centering
\caption{Slater-Usoh-Steed Presence Questionnaire Results.}\label{tab:respres}
\scriptsize
\begin{tabular}{lrrrrrrr}\toprule
\multirow{2}{*}{\textbf{Participants}} &\multicolumn{6}{c}{\textbf{Questions}} \\\cmidrule{2-7}
&\textbf{Q1} &\textbf{Q2} &\textbf{Q3} &\textbf{Q4} &\textbf{Q5} &\textbf{Q6} \\\midrule
\textbf{Mean} &\textbf{5.5} &\textbf{5.8} &\textbf{3.5} &\textbf{4.3} &\textbf{4.6} &\textbf{5} \\
\textbf{Std Dev} &\textbf{1.04} &\textbf{1.16} &\textbf{2.81} &\textbf{2.8} &\textbf{2.16} &\textbf{1.67} \\
\bottomrule
\end{tabular}
\end{table}

%  \begin{figure}[htb]
%  \centering% avoid the use of \begin{center}...\end{center} and use \centering instead (more compact)
%  \includegraphics[width=\linewidth]{pictures/presence.png}
%  \caption{Slater-Usoh-Steed Presence Questionnaire Mean Rating Per Question.}
%  \label{fig:pres}
% \end{figure}

\subsection{Overall Perceptions}

On average all questions in the Overall Perceptions questionnaire responses were above 3.5 in a scale of 1 to 5 where 5 is best.

% \begin{figure}[htb]
%  \centering% avoid the use of \begin{center}...\end{center} and use \centering instead (more compact)
%  \includegraphics[width=\linewidth]{pictures/perceptions.png}
%  \caption{Overall Perceptions Questionnaire Mean Ratings.}
%  \label{fig:perceptions}
% \end{figure}

\section{Discussion}

The results of the usability study indicate that the SURVIVRS prototype is a usable tool that elicits a high sense of presence and is perceived as effective and sufficient by non-medical users.

% Recall the research questions and hypotheses:
% \begin{quote}
% RQ1. What is the overall usability of the VR software?

% ~~~H1. The interface will be regarded as having a high degree of usability.

% RQ2. Are the VR user interface tools perceived as sufficient and helpful for the users to offer remote guidance to a surgery team?

% ~~~H2. Users will perceive the VR user interface tools as sufficient and helpful

% RQ3.  Does the experience induce a sense of presence?

% ~~~H3. Users will feel a high degree of presence during the VR experience.
    
% \end{quote}

%Prior research found that the overall score of the System Usability Scale can be considered average at a score of 68\footnote{https://www.usability.gov/how-to-and-tools/methods/system-usability-scale.html}.
%The results of the overall System Usability Scale score indicate the usability of the User Interface (UI) was above the average for most of the participants. Besides, SUS mean rating per question is, on average equal to 4.5 out of 5. From Slater-Usoh-Steed, all the users were above the average in the results, which means the presence of being inside a virtual environment was good for the recorded video mode in the application.

%We faced two types of implications in this experiment: experimental issues where the evaluation was done with seven participants but reported results for six participants only because one participant didn’t take it seriously. And in the interface issues, there were many notes from the experimenter while observing the participants during the tasks. 
Several points were noted during the experiment. For example, five participants had difficulty drawing in their first attempt, but drawing became easier the second time. Also, no participant understood that the play/pause button also took screenshots. Furthermore, some participants wondered why zooming out screenshots has a different way than zooming out the videos. This makes a bad user experience (UX) impression. Also, four out of seven participants, while trying to click on the patient monitor video to zoom it in, by mistake, clicked on the home button and got out of the application.  Other comments included slight dizziness (1), need for help (2), slanted 360\degree~video (3), feeling small (1), and realistic hospital feeling, but lacking hospital odors (1).

%; during the task, some participants kept trying/clicking on all buttons until they found it out, but others couldn’t complete taking a screenshot task. All of them didn’t understand the logic behind making the play/pause button take screenshots.

% The participants in the questionnaire put some important notes to be considered; one of them saw that the videos were very responsive to his head movements, and he felt some discomfort and dizziness. Two other participants express their need for help or direct instructions when using this application. Although four participants found the surgery room from the video very similar to a real place, others declare that the video shape makes it look slanted. 

% Even though the results from the questionnaires about the sense of presence in VE results were acceptable, one participant commented that he/she felt small in their surroundings, and another said he felt like he was in the hospital, but nothing smelled like a real hospital. 

% Some recommendations to consider enhancing or fixing the user interface of the vr application: the first one is about fixing/changing the way of taking screenshots, may keep the play/pause button be able to take screenshots easier for the surgeon, but may add another button for screenshots only will improve the usability of ui. Second, change the position of the home button. Third, deal with the zooming in/out by using the same way and buttons for videos and screenshot to enhance the level of ux.	 

\section{Conclusion and Future Work}
\label{sec:conclusion}

We proposed SURVIVRS, a surgery guidance tool that allows remote expert surgeons to guide medical teams without high expertise. Our prototype tool showed positive results, and is a promising first step towards realizing the full vision of SURVIVRS.

There are several next steps to advance our project. After feedback-based refinements are made, we plan on testing the user interface with surgeons to gain domain-specific feedback. We also need to test the tool with a live stream of a (simulated) surgery and address the many technical challenges associated with it. At the medium term, we will develop a local surgeon interface using monitors in the surgery room or smart glasses.

In the long term, we aim to incorporate augmented reality tools in the surgery room to close the telepresence loop and integrate a representation of the remote surgeon and spatial in-context annotations.

\acknowledgments{
This work is being partially supported by \textit{Fundação para a Ciência e a
Tecnologia} (Portuguese Foundation for Science and Technology) through
grants 2022.09212.PTDC (XAVIER), UIDB/50021/2020 and Carnegie Mellon
Portugal grant SFRH/BD/151465/2021 under the auspices of the UNESCO
Chair on AI\&XR.}

\balance

\bibliographystyle{abbrv-doi}

\bibliography{template}

\end{document}